\documentclass[pdflatex, sn-mathphys-num, iicol]{sn-jnl}

\usepackage{graphicx}%
\usepackage{multirow}%
\usepackage{amsmath,amssymb,amsfonts}%
\usepackage{amsthm}%
\usepackage{mathrsfs}%
\usepackage[title]{appendix}%
\usepackage{xcolor}%
\usepackage{textcomp}%
\usepackage{manyfoot}%
\usepackage{booktabs}%
\usepackage{algorithm}%
\usepackage{algorithmicx}%
\usepackage{algpseudocode}%
\usepackage{listings}%

\theoremstyle{thmstyleone}%
\theoremstyle{thmstyletwo}%

\theoremstyle{thmstylethree}%

\usepackage{xcolor}

\begin{document}

\title[Ultrafast formation of a large dynamic magnetic soliton]{Ultrafast formation of a large dynamic magnetic soliton}

\author*[1,2]{\fnm{Ondřej} \sur{Wojewoda}}\email{ondrej.wojewoda@vutbr.cz}
\author[3,4]{\fnm{Sina} \sur{Mayr}}
\author[5]{\fnm{Miela J.} \sur{Gross}}
\author[2]{\fnm{Jan} \sur{Klíma}}
\author[2]{\fnm{Jaganandha} \sur{Panda}}
\author[2]{\fnm{Jakub} \sur{Krčma}}
\author[2]{\fnm{Jakub} \sur{Holobrádek}}
\author[6]{\fnm{Kristýna} \sur{Davídková}}
\author[6]{\fnm{Andrii V.} \sur{Chumak}}
\author[7]{\fnm{Igor} \sur{Gerasimchuk}}
\author[7]{\fnm{Roman} \sur{Verba}}
\author[9]{\fnm{Philipp} \sur{Pirro}}

\author[8]{\fnm{Markus} \sur{Weigand}}
\author[3]{\fnm{Simone} \sur{Finizio}}

\author[10]{\fnm{Morris} \sur{Lindner}}
\author[10]{\fnm{Carsten} \sur{Dubs}}

\author[11]{\fnm{Qi} \sur{Wang}}

\author[8]{\fnm{Sebastian} \sur{Wintz}}
\author[1]{\fnm{Caroline A.} \sur{Ross}}
\author[2]{\fnm{Michal} \sur{Urbánek}}

\affil[1]{\orgdiv{Department of Materials Science and Engineering}, \orgname{Massachusetts Institute of Technology}, \orgaddress{\city{Cambridge}, \state{MA}, \country{USA}}}
\affil[2]{\orgdiv{CEITEC BUT}, \orgname{Brno University of Technology}, \orgaddress{ \city{Brno}, \country{Czech Republic}}}
\affil[3]{\orgname{Paul Scherrer Institut}, \orgaddress{ \city{Villigen}, \country{Switzerland}}}
\affil[4]{\orgname{Laboratory for Mesoscopic Systems, Department of Materials, ETH}, \orgaddress{ \city{Zürich}, \country{Switzerland}}}
\affil[5]{\orgdiv{Department of Electrical Engineering and Computer Science}, \orgname{Massachusetts Institute of Technology}, \orgaddress{\city{Cambridge}, \state{MA}, \country{USA}}}
\affil[6]{\orgdiv{Faculty of Physics}, \orgname{University of Vienna}, \orgaddress{ \city{Vienna}, \country{Austria}}}
\affil[7]{\orgname{V. G. Baryakhtar Institute of Magnetism of the NAS of Ukraine}, \orgaddress{ \city{Kyiv}, \country{Ukraine}}}
\affil[8]{\orgname{Helmholtz-Zentrum-Berlin für Materialien und Energie}, \orgaddress{ \city{Berlin}, \country{Germany}}}
\affil[9]{\orgdiv{Fachbereich Physik and Landesforschungszentrum OPTIMAS}, \orgname{Technische Universität
Kaiserslautern}, \orgaddress{ \city{Kaiserslautern}, \country{Germany}}}
\affil[10]{\orgname{INNOVENT e.V. Technologieentwicklung}, \orgaddress{ \city{Jena}, \country{Germany}}}
\affil[11]{\orgdiv{School of Physics}, \orgname{Huazhong University of Science and Technology}, \orgaddress{ \city{Wuhan}, \country{China}}}

\abstract{Nonlinear magnetization dynamics offers a rich variety of phenomena ranging from bistability to chaos. Here, we report the ultrafast formation of a dynamic magnetic soliton in thin ferrimagnetic garnet films with perpendicular magnetic anisotropy, driven by the microwave magnetic field of a microstrip antenna. Using time-resolved Brillouin light scattering microscopy and scanning transmission X-ray microscopy, we directly track the build-up of the large-angle precession state. The observed soliton is distinct from other nonlinear magnetic excitations in two key aspects: (i) it forms inside the linear spin-wave frequency band, and (ii) it is exceptionally large, reaching tens of microns beyond the antenna. We explain the soliton formation by the self-limiting mechanism upon a positive nonlinear frequency shift and the spatial extent of the near-field of the antenna. At large distances from the drive, the soliton collapses and emits short-wavelength spin waves via almost instantaneous spatial wavenumber conversion. Time-resolved measurements further reveal a small finite delay during soliton formation, while coherent long-range oscillations appear essentially simultaneously over distances up to 40\,\textmu m. These results establish microwave-driven solitons as a robust nonlinear phenomenon in thin-film garnets and suggest opportunities for fast, nonlocal manipulation of magnetic states and for applications in novel computational schemes.}

\keywords{Ultrafast, magnon, spin-wave, magnetic soliton, nonlinear physics}

\maketitle
\noindent\parindent=1em 
\begin{figure*}
\centering
\includegraphics[width=1\textwidth]{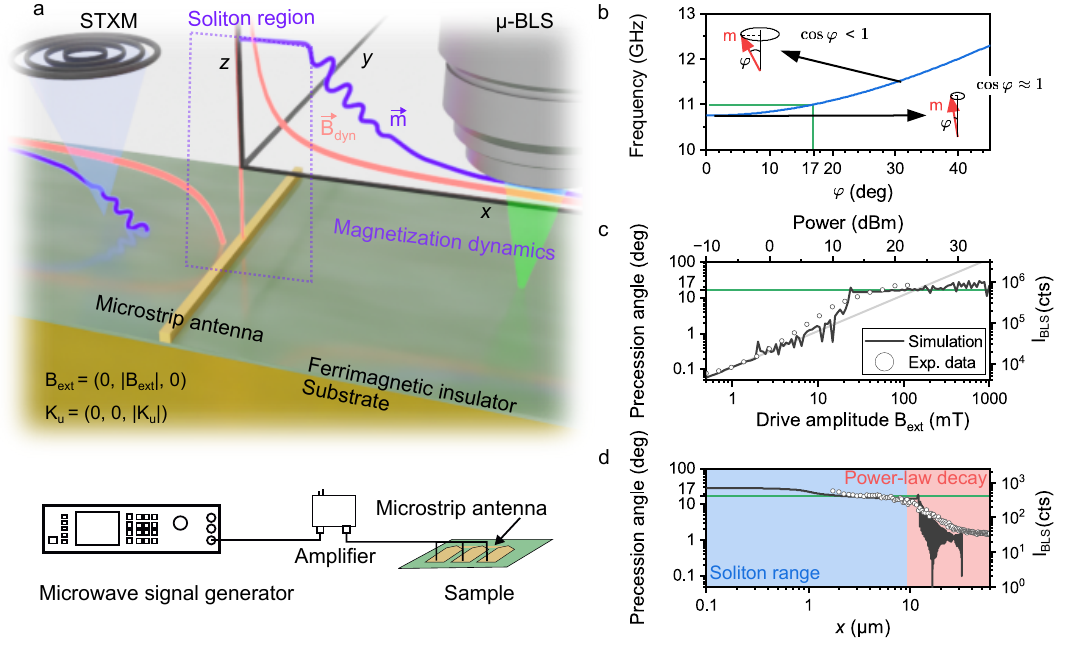}
\caption{\textbf{Schematic picture of the soliton formation and spin-wave emission}. \textbf{a} Experimental geometry. The excitation stripline is driven by an RF (radio frequency) generator with a single-frequency signal. The resulting magnetization dynamics (purple line) is subsequently measured by µ-BLS with intensity, time, and phase resolution or by time-resolved STXM. \textbf{b} Nonlinear shift of the ferromagnetic resonance as a function of the precession angle ($\varphi$). The inset illustrates an increase in the precession amplitude. The excitation frequency of 11\,GHz is marked by the green line. Calculations use Bi:YIG material parameters (see Methods) and an in-plane external magnetic field of 550\,mT. \textbf{c} Precession angle ($\varphi$) as a function of the driving field amplitude. The upper axis shows the applied microwave power and is scaled to overlap the experimental data with the simulated curve. \textbf{d} Spatial decay of the precession angle ($\varphi$) with distance from the microstrip antenna.}\label{fig1}
\end{figure*}

Computational functionality requires a physical system capable of a nonlinear response. Historically, such nonlinearity has been provided by transistors, which exhibit threshold behavior and enable the realization of logic gates and modern computer architectures. Recently, computational paradigms themselves are being rethought, with novel approaches emerging, including inverse-designed devices and large-scale neural networks \cite{lecun2015deep, Wang2021, Shalf2020}. However, implementing these architectures within existing hardware platforms remains challenging. To address this issue, a range of alternative physical systems is currently being explored. Among them, magnetization dynamics stands out, as it provides intrinsic nonlinearity at experimentally accessible power levels, enables electrical readout, and can directly interface with long-term magnetic information storage \cite{baumgaertl2023reversal, fan2023coherent, bader2010spintronics, hirohata2020review}.

Magnetization dynamics is intrinsically nonlinear because the magnetization vector is constrained to move on a spherical phase space. This gives rise to a broad spectrum of nonlinear phenomena, including foldover, bistability, multi-magnon scattering, chaos, and soliton formation \cite{Wigen_Book1994, Bertotti_Book, pirro2021advances, chumak2015magnon, Dumas2014, wang2023deeply, Turenne2022, bonetti2015direct, Rajabali2023, ustinov2021, Rezende1990}. Among these phenomena, dynamic solitons -- spatially localized, aperiodic in space nonlinear excitations -- are particularly attractive because their large precessional amplitudes provide a route to efficient manipulation of magnetic states. Dynamic solitons generated by microwave pulses or spin-transfer torque (STT) have been studied extensively. However, most previous work has focused either on conservative exchange solitons \cite{Kosevich_Physica1981, Kosevich1990} or on STT-driven dissipative solitons \cite{Slavin_PRL2005, Sulymenko2018, Macia2020, Backes2015}. These excitations are typically confined to length scales of only a few exchange lengths, or remain localized near the active STT region, which limits their usefulness for information transfer. Dipolar interactions can relax this strong localization and support more extended dissipative solitons \cite{Ivanov_JETPLett1979, Verba_LTP2020}, but only under specific conditions. Conversely, propagating spin-wave envelope solitons are usually very large, extending over tens to hundreds of spin-wave wavelengths, and exhibit comparatively small amplitudes \cite{Gasperis1988, Kovsikov_PRB1996, Serga2005, Sulymenko2018}. The regime most relevant for magnonic circuitry -- a nonlinear excitation formed near an RF-pumped microstrip antenna, with a spatial extent and amplitude sufficient to manipulate a magnonic gate -- has so far remained largely unexplored.

Here, we present time-resolved Brillouin light scattering microscopy (BLS) and scanning transmission X-ray microscopy (STXM) measurements of the formation of a nonlinear localized excitation -- a dynamic dissipative soliton, which forms above the linear spin-wave frequency onset, in sharp contrast with conservative and common dissipative solitons. The soliton is stabilized by the microwave magnetic field of the antenna, which reaches amplitudes of up to 15\,mT, enabling the emergence of nonlinear dynamics \cite{wang2023deeply}. This nonlinearly stabilized soliton exhibits an exceptionally large spatial extent of tens of microns. Beyond the soliton localization area, which is determined by the strength and frequency of the driving field, the nonlinear localized state collapses and spin waves are emitted from the boundary region.

We have demonstrated soliton formation in two thin-film garnet samples with perpendicular magnetic anisotropy, Bi-substituted yttrium iron garnet (Bi:YIG) epitaxially grown on gadolinium scandium gallium garnet (GSGG) \cite{fan2023coherent, fakhrul2019magneto} and Ga-substituted yttrium iron garnet (Ga:YIG) epitaxially grown on gadolinium gallium garnet (GGG) \cite{Dubs2025, Botcher2022, Carmiggelt2021, Breitbach2024nonlinear}. The Bi:YIG was investigated by BLS and the Ga:YIG by STXM in an in-plane field (Fig.~\ref{fig1}a); for more details, see the Methods section and Extended Data Fig.~\ref{SI-Char},~\ref{SI-GaYIGChar}.

\section*{Formation of the soliton}\label{sec1}
The existence and properties of dynamic magnetic solitons are controlled primarily by the nonlinear frequency shift and the magnon dispersion relation. In both systems investigated here, the applied in-plane magnetic field is sufficient to saturate the magnetization along the field direction. In this regime, the nonlinear frequency shift of the ferromagnetic resonance (FMR) and of long-wavelength, low-$k$ spin waves is positive, owing to the perpendicular magnetic anisotropy field exceeding the demagnetization field.

We investigate the Bi:YIG sample under excitation by an external microwave magnetic field at a fixed frequency of 11\,GHz generated by the antenna. This frequency lies above the linear FMR frequency, as indicated by the green solid line in Fig.~\ref{fig1}b. As shown in Extended Data Fig.~\ref{S-Dispersion}, the linear spin-wave dispersion in substituted YIG increases approximately quadratically with wavenumber because of the dominant exchange contribution. The excitation frequency therefore lies within the spin-wave band and corresponds to a high-$k$ mode with a wavelength of approximately 500~nm, which is not directly accessible by the micron-sized antenna. At low driving fields, the magnetization is therefore driven far off the FMR and the precession amplitude increases linearly with the applied microwave field (Fig.~\ref{fig1}c). As the driving field is increased, the precession angle grows until the nonlinear frequency shift brings the FMR into resonance with the excitation. Under our conditions, this occurs at a precession angle of approx. 17$^\circ$ (Fig.~\ref{fig1}c). Further increase of the driving field does not lead to a higher precession amplitude: instead, the precession angle becomes locked at this value, as any additional increase in amplitude shifts the resonance to higher frequencies and thereby reduces the coupling efficiency to the fixed-frequency drive \cite{wang2023deeply}. This self-limiting mechanism stabilizes the precession angle and renders its amplitude largely independent of further increases in the excitation power.

If the driving field were strictly confined under the antenna, the nonlinear quasi-uniform FMR state would be limited to the antenna region, beyond which the magnetization dynamics would immediately evolve into propagating spin waves \cite{wang2023deeply}. This conclusion follows from the analytical model, see Methods, and is supported by a numerical solution (Extended Data Fig.~\ref{S-th}b). A real microstrip antenna, however, exhibits spatially extended near fields, which decrease with distance from the antenna, scaling approximately as $1/x$, and these fields can sustain nonlinear precession beyond the antenna region. In a simplified qualitative picture, quasi-uniform nonlinear precession remains stable until the antenna field eventually drops below the threshold value required to sustain local nonlinear FMR. At this point, the locked precession angle collapses, and the BLS signal decreases with power-law attenuation governed by the spatial decay of the driving field from the antenna.

\begin{figure*}[!ht]
\centering
\includegraphics[width=1\linewidth]{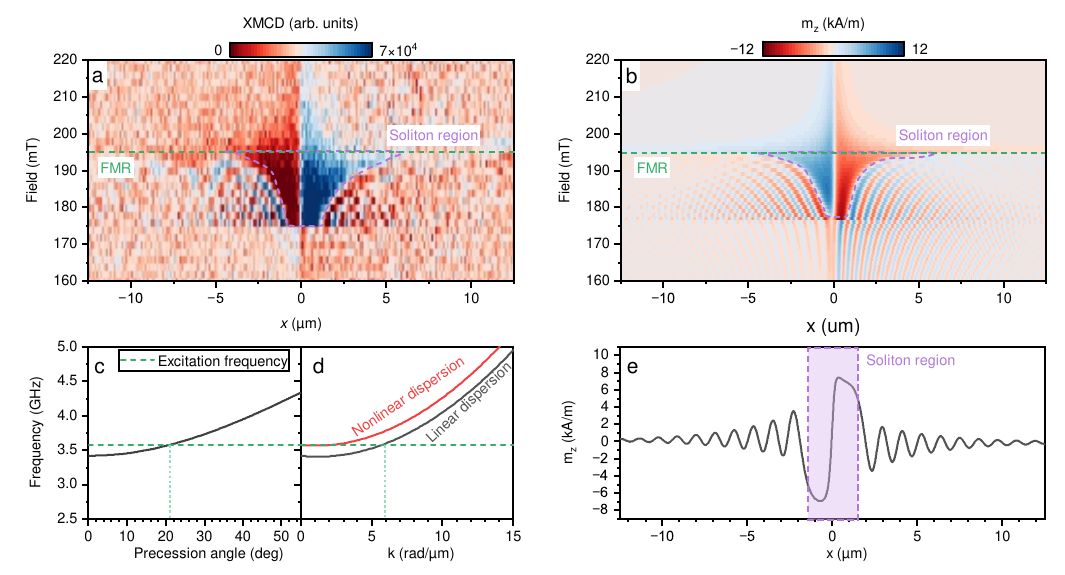}
\caption{\textbf{Emission of spin waves at the soliton boundary in Ga:YIG}. \textbf{a, b} Snapshots from time-resolved STXM \textbf{(a)} and simulation \textbf{(b)} at a fixed driving frequency of 3.57\,GHz while sweeping the external magnetic field. \textbf{c, d} Nonlinear FMR shift at a precession angle of approx. 20$^\circ$ \textbf{(c)} and the corresponding linear and nonlinear dispersion relations \textbf{(d)} for the Ga:YIG sample at an external field of 190\,mT. The green dashed line shows the driving field frequency of 3.57\,GHz. Dashed lines show the expected precession angle and wavenumber. \textbf{e,} Line profiles of the simulated intensity in \textbf{(b)} at 190\,mT.}\label{fig4}
\end{figure*}

The real physical picture of the magnetization dynamics is more complex and goes beyond a simple local nonlinear FMR description; in particular, magnon group-velocity dispersion plays an important role. Because the resulting nonlinear excitation is spatially localized and aperiodic, we refer to it as a dynamic soliton. Since it cannot exist without continuous pumping, it is a dissipative soliton. Its profile is determined by the intrinsic nonlinear mechanisms and does not simply follow the driving-field profile (Extended Data Fig.~\ref{S-th}a). Moreover, the spatial extent of the soliton is set by the driving-field strength and frequency and can be tuned over a wide range (Fig.~\ref{fig4}, Extended Data Fig.~\ref{S-th} and \ref{S-PowerFreq}).

\section*{Emission of spin waves at the soliton boundary}\label{sec3}
The Bi:YIG film of Fig. 1 is only 21\,nm thick, limiting the decay length of propagating spin waves
to below $\approx$1\,µm. To access the magnetization dynamics at the soliton boundary, we turn to a 161-nm-thick Ga:YIG film analyzed via STXM, which provides higher spatial resolution. Moreover, owing to the reduced saturation magnetization of Ga:YIG, the group velocity of low-$k$ spin waves, which are also exchange-dominated, becomes higher, increasing the spin-wave decay length to several micrometers.

We excite the Ga:YIG sample at a fixed frequency of 3.57\,GHz in the Damon-Eshbach-like geometry while sweeping the external magnetic field (Fig.~\ref{fig4}a). The same experimental configuration was also simulated using MuMax3 (Fig.~\ref{fig4}b). In both panels, we observe that at high fields, when the driving field frequency is below the FMR, no signal is detected except for a small region of forced oscillation in the immediate vicinity of the antenna. At a field of $195\,$mT, the FMR is excited, with opposite phase at the two ends of the antenna, reflecting the spatial structure of the driving field \cite{Schneider2008}. When we decrease the field further, a localized p-like soliton forms, and its width progressively shrinks with decreasing field. This trend is consistent with our analysis of Bi:YIG (Fig.~\ref{fig1}). After the field decreases below $\approx 175\,$mT, the soliton collapses and no further excitation is visible.

In the following, we focus on the field value $B_\mathrm{ext}=190$\,mT, at which the driving field frequency lies above the linear FMR. The resonance condition at 3.57\,GHz can therefore only be met via the nonlinear upshift in frequency (Fig.~\ref{fig4}c). From the required frequency shift, we infer a precession angle of approx. 20$^\circ$ and use it to calculate the dispersion relation (Fig.~\ref{fig4}d) \cite{klima2026swt}. Strikingly, once the soliton collapses at the boundary, where the local microwave field drops below the threshold needed to sustain the required large-angle precession, it emits propagating short-wavelength spin waves (Fig.~\ref{fig4}e). The transition from nonlinear quasi-uniform precession to propagating spin waves occurs almost immediately in space (the transition region cannot be resolved in Fig.~\ref{fig4}b,e). This is similar to the case of step-like microwave excitation \cite{wang2023deeply}.

The emission of spin waves is also visible in the experimental data measured on the Ga:YIG sample (Fig.~\ref{fig4}a) and shows the same features as the simulations. In contrast, in the Bi:YIG sample, we were unable to detect spin-wave emission because of the faster decay of the emitted waves and their overlap with weak forced long-range oscillations excited by the spatially decaying antenna field \cite{Miela2026}.

This mechanism of short-wavelength spin-wave emission can be understood as wavenumber conversion enforced by the boundary between a strongly nonlinear region (large precession angle and strongly upshifted dispersion; red curve in Fig.~\ref{fig4}c) and a linear region (small precession angle and linear dispersion; black curve in Fig.~\ref{fig4}c). In the nonlinear region beneath and near the antenna, the dynamics is dominated by a quasi-uniform mode (effectively $k\to0$), whose frequency is pulled upward by the large precession angle. Outside the soliton region, the oscillation amplitude decreases, the dispersion rapidly shifts back toward the quasi-linear branch, and frequency conservation forces the excitation to transfer to substantially larger wavenumbers, i.e., to shorter wavelengths. The nonlinearly stabilized solitons and short-wavelength spin-wave emission can also be observed in out-of-plane magnetized layers, but the solitons are confined to below the antenna as the out-of-plane dynamic field cannot exert torque on the magnetization and hence cannot drive spatially-extended solitons \cite{wang2023deeply}.

\begin{figure*}
\centering
\includegraphics[width=1\textwidth]{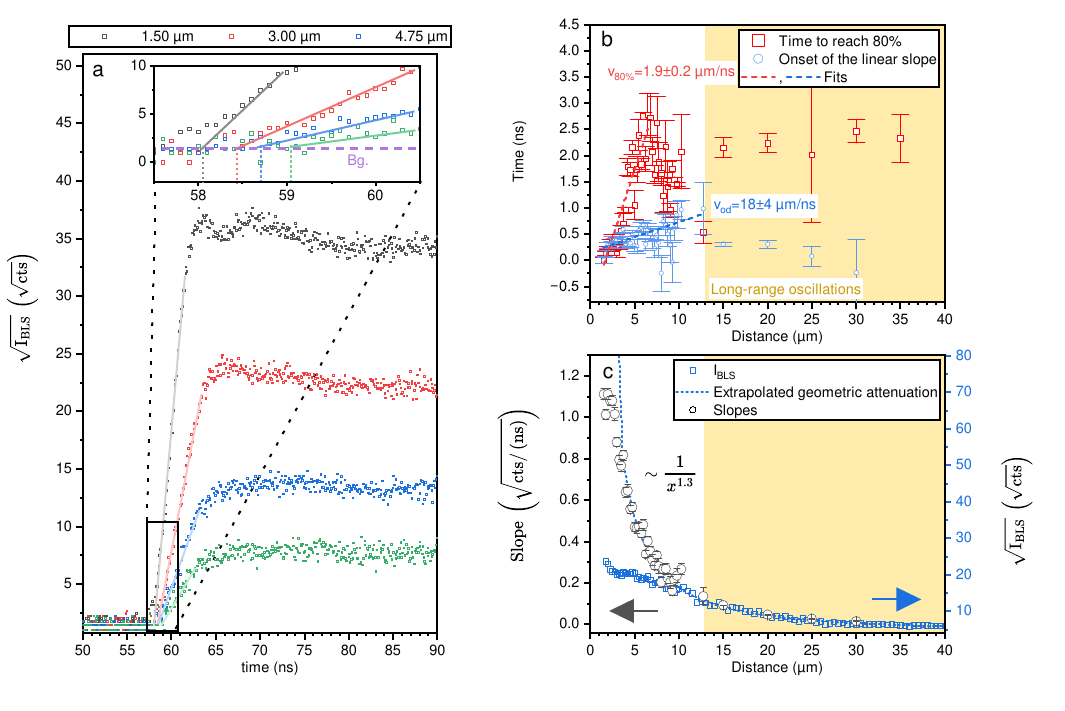}
\caption{\textbf{Time-resolved observation of the soliton formation and long-range oscillations in Bi:YIG.} \textbf{a,} Time traces of the square root of the BLS intensity recorded at different distances $x$ from the antenna. Linear fits to the rising edges are depicted as faded solid lines. The inset highlights the distance-dependent onset delay of the soliton.\textbf{b,} Times of arrival extracted from the time to reach 80\,\% (red circles) and a linear fit used to determine the onset delay velocity in the nonlinear region (slope). In the linear region, the long-range oscillations follow the antenna excitation without a measurable delay. \textbf{c,} Slopes of the rising-edge fits from panel \textbf{(a)} (black circles). Blue squares represent the square root of the BLS signal versus distance. }\label{fig3}
\end{figure*}

\section*{Time-resolved measurements of the soliton formation}\label{sec2}

To resolve the transient dynamics of soliton formation, we performed pulsed time-resolved µ-BLS on the Bi:YIG sample. We recorded the BLS intensity as a function of time at different positions relative to the antenna (Fig.~\ref{fig3}a).

At the position closest to the antenna ($x = 1.5\,$µm), the BLS signal exhibits pronounced oscillations immediately after the rising edge (Fig.~\ref{fig3}a, black symbols). We attribute these oscillations to beating between forced and free-running magnetization dynamics, which commonly occurs when an oscillator is driven off resonance by an abruptly switched-on force. In the present case, the driving field frequency differs from the linear FMR frequency, while the large drive amplitude and strong nonlinearity further enhance the beating response. As the precession amplitude builds up, the nonlinear frequency progressively upshifts the effective FMR toward the drive frequency. Consequently, the beating frequency decreases with time and ultimately vanishes once the resonance condition is reached and the system settles into a stable nonlinear state, i.e., the oscillation becomes phase-locked to the drive as the nonlinear shift compensates the initial detuning. Farther from the antenna ($x=3; 4.75\,$µm), the beating is no longer visible (Fig.~\ref{fig3}a, blue symbols), as nonlinear effects become weaker.

The signal builds up more slowly with increasing distance from the antenna, which can be understood as a consequence of the reduced driving field. To quantify this effect, we fit the rising edge using the linear dependence $\sqrt{I_{\mathrm{BLS}}} \approx (t-t_0)$. Such behavior is expected for a simple damped harmonic oscillator (see Methods) and should provide a reasonable approximation at the earliest stage of magnetization precession, when the amplitude remains too small for nonlinear effects to become significant. The measured rising edge is approximately linear, allowing us to estimate the onset time of the oscillations from the intersection of the linear fit with the background level (see inset in Fig.~\ref{fig3}a). The extracted onset times exhibit a distance-dependent delay. A linear fit to these onset delay times yields an apparent velocity of $v_\mathrm{od} = (18\pm4)\,$µm\,ns$^{-1}$ (Fig.~\ref{fig3}b). We emphasize that this delay should not be interpreted as the group velocity of propagating spin waves. Instead, it likely reflects nontrivial phase-evolution processes during soliton formation, influenced by the thermally excited magnon background (Extended Data Fig.~\ref{S-PR-TR}). The magnetization dynamics are expected to start simultaneously across the driven region; however, the early evolution remains hidden below the thermal level. This apparent delay is therefore not a propagation delay in the conventional sense, but it remains relevant for applications.

In addition, because the amplitude is nonlinearly limited and the slope of the rising edge depends on the driving-field amplitude, the time at which the oscillations approach their steady state also varies spatially. To capture this behavior, we analyzed the time required for the amplitude to reach 80\,\% of its steady-state value (Fig.~\ref{fig3}b). A linear fit to these delay times yields an effective velocity of $v_\mathrm{80\%}=(1.9\pm0.2)\,$µm\,ns$^{-1}$. The time required to form a high-amplitude soliton is orders of magnitude shorter than that required for linear spin waves to reach the same distances (spin-wave group velocities are on the order of tenths of µm~ns$^{-1}$, see Extended Data Fig.~\ref{S-Dispersion}). This suggests a pathway toward faster magnetization manipulation~\cite{caretta2020relativistic}. At larger distances, where the driving field is insufficient to form a soliton, no additional delay is observed; instead, the long-range oscillations appear essentially instantaneously because the nonlinear mechanisms are no longer active. This is further corroborated in the Ga:YIG sample by pulsed STXM measurements (Extended Data Fig.~\ref{S-GaYIG-XMCD}).

In the linear regime, the magnetization amplitude should scale proportionally with the driving field, as observed at larger distances ($x>13\,$µm, blue squares in Fig.~\ref{fig3}c). A fit at larger distances yields a power-law attenuation of the rising edge slope proportional to $\approx 1/x^p$ with $p = 1.3 \pm 0.1$. For an ideal, infinitely long wire one would expect $p=1$; the deviation can be attributed to the finite length (20\,µm) of the real antenna and to the presence of the second branch of the antenna carrying current in the opposite direction. Extrapolating the fitted attenuation rule closer to the antenna yields good agreement with the measured slopes, consistent with their proportionality to the driving field amplitude (black circles in Fig.~\ref{fig3}c). In contrast, $\sqrt{I_\mathrm{BLS}}$ saturates near the antenna and deviates from the field-defined scaling rule, reflecting the onset of the nonlinear response.

\begin{figure}
\centering
\includegraphics[width=1\linewidth]{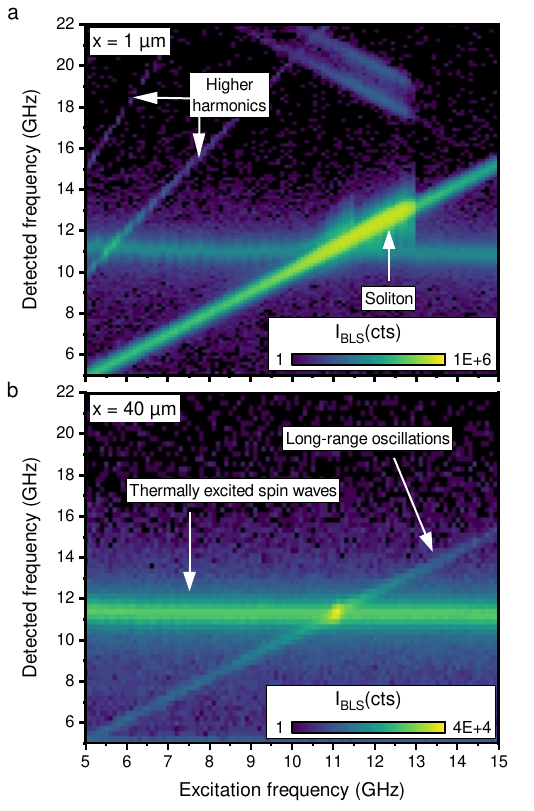}
\caption{\textbf{Frequency sweeps near and far from the antenna in Bi:YIG}. \textbf{a, b} µ-BLS measurement of magnetization dynamics at $x=1\,$µm \textbf{(a)} and $x=40\,$µm \textbf{(b)}. The acquisition time for panel \textbf{(b)} was significantly longer. In panel \textbf{(a)} we used pulsed excitation to limit the heating effects. The color code is logarithmic. The signal with negative slope is caused by the high signal and higher-order transmission through the interferometer.}\label{fig2}
\end{figure}

\section*{Nonlinear spectral features and long-range forced response}

We performed frequency sweeps at two fixed distances from the antenna (Fig.~\ref{fig2}a,b). Near the antenna ($x = 1\,$µm) and at low excitation frequencies, we observe an upshift of the thermally excited spin-wave background, which we attribute to the local heating of the antenna. The ripple-like modulation originates from the frequency-dependent microwave power transmission through the antenna. In addition, we detect weak higher-order signals at the second and third harmonics of the drive (2$f$ and 3$f$). These harmonics were not observed in the micromagnetic simulations, suggesting that they either correspond to multi-magnon scattering enabled by sample defects or magnetic inhomogeneities, or are caused by the presence of higher harmonics in the excitation signal. In any case, higher harmonic amplitudes remain very low, and they do not significantly affect the phenomena discussed above. Once the applied power is no longer sufficient to drive the magnetization with the required precession angle (approximately $50^\circ$ at 13\,GHz), the system moves out of resonance, and the overall BLS signal is reduced by several orders of magnitude.

At the larger distance ($x = 40\,$µm), all nonlinear phenomena disappear. The spectrum is then dominated by the thermally excited spin waves (which appear as a frequency-independent band around 11\,GHz) and by the coherently excited long-range oscillations. At this distance and applied power of 22\,dBm, we estimate the local driving-field amplitude to be $\approx 0.3\,$mT. The driven long-range oscillations are present across the entire frequency range, with a pronounced amplitude maximum near the FMR. They exhibit no measurable group delay and thus appear simultaneously at all investigated frequencies, which is consistent with a forced response to the antenna field rather than propagating spin waves. Owing to the strong magneto-optical response of Bi:YIG, these oscillations remain detectable even far from resonance; however, their amplitude is very low. Near resonance, they can reach considerable precession angles (up to $\approx 0.1\,^\circ$), enabling immediate excitation of additional magnetization dynamics, such as domain-wall motion \cite{fan2023coherent}.

\section*{Conclusion}\label{sec4}
In conclusion, we demonstrate the formation of a dynamic dissipative magnetic soliton at a frequency within the linear spin-wave band in ferrimagnetic garnet films with perpendicular magnetic anisotropy under strong microwave pumping. Using time-resolved µ-BLS (Bi:YIG) and time-resolved STXM (Ga:YIG), we directly tracked the spatiotemporal build-up of the large-angle precession state and its evolution with drive strength and magnetic field. The soliton with unprecedentedly large size, reaching tens of micrometers, is stabilized by a nonlinear self-limiting mechanism under a positive nonlinear frequency shift and by the near fields of the antenna: the nonlinear frequency shift pulls the local FMR into resonance with a fixed-frequency drive and locks the precession amplitude, resulting in a spatially confined excitation with a tunable extent. At a certain distance, defined by the local driving field, excitation frequency, and magnon group-velocity dispersion, the soliton collapses and emits short-wavelength spin waves via wavenumber conversion at the boundary between the strongly nonlinear region and the surrounding linear region. This emission is observed in Ga:YIG, where the spin-wave decay lengths extend to several micrometers, whereas in the thinner Bi:YIG film the emitted waves are quickly attenuated and the response transitions primarily to the long-range weak forced oscillations.
Time-resolved measurements further reveal a finite, distance-dependent delay during soliton formation; the corresponding effective group velocities exceed those expected for linear spin waves. In contrast, long-range oscillations far from the antenna follow the microwave drive without a measurable delay over distances up to 40\,µm, which is consistent with a forced response rather than spin wave propagation. These findings establish antenna-driven dissipative solitons with ultrafast formation dynamics as controllable nonlinear elements for magnonic signal processing. Their amplitude self-limiting and boundary-mediated wavenumber conversion provide mechanisms for thresholding, gating, and activation-like functions in neuromorphic or inverse-designed magnonic devices, as well as passive RF components such as power limiters and reconfigurable filters \cite{devitt2026spin}.

\backmatter

\bmhead{Data and code availability}
Data and simulation codes are available from the Zenodo repository \cite{zenodo}.


\bmhead{Acknowledgements}
C.D. thanks Oleksii Surzhenko for the VSM measurements and R. Meyer for technical support. The authors acknowledge the MIT Office of Research Computing and Data for providing high-performance computing resources and thank Helmholtz-Zentrum Berlin for the allocation of synchrotron radiation beamtime. We acknowledge CzechNanoLab Research Infrastructure (ID 90251), supported by MEYS CR. O.W. was supported by Horizon Europe - MSCA grant agreement No. 101211677, project FeriMag. S.M. was funded by the Swiss National Science Foundation (Grant Agreement No. 172517). J.Kl. was supported by the Brno Ph.D. Talent scholarship issued by Brno City Municipality. C.D. acknowledges the Deutsche Forschungsgemeinschaft (DFG, German Research Foundation) for funding project 271741898. I.G. and R.V. acknowledge support by the National Research Foundation of Ukraine, Grant No. 2025.07/0132. C.R. acknowledges support of NSF award ECCS 2423929. This research was supported by the Project No. CZ.02.01.01/00/22 008/0004594 (TERAFIT).

\bmhead{Author contributions}
Conceptualization: O. W., R. V. and M. U.; Data curation: O. W.; Formal analysis: O. W., R. V., C. D., Q. W., S. W., C. A. R. and M. U.; Funding acquisition: O. W. and M. U.; Investigation: O. W., S. M., M. J. G., J. H., I. G., R. V., M. W., S. F., M. L., C. D. and S. W.; Methodology: O. W., S. M., M. W. and S. F.; Resources: M. J. G., J. P., J. H., K. D., M. L. and C. D.; Software: O. W., J. Kl., J. Kr. and R. V.; Supervision: A. V. C., R. V., P. P., Q. W., S. W., C. A. R. and M. U.; Validation: O. W.; Visualization: O. W. and R. V.; Writing - Original Draft: O. W.; Writing - Review \& Editing: S. M., M. J. G., J. Kl., J. P., J. Kr., J. H., K. D., A. V. C., I. G., R. V., P. P., M. W., S. F., M. L., C. D., Q. W., S. W., C. A. R. and M. U.;
\bmhead{Conflict of interest}
The authors declare no competing interests.




\section*{Methods}
\subsection*{Characterization of the samples}
The single-crystal quality of the epitaxial Bi:YIG (Bi$_{0.8}$Y$_{2.2}$Fe$_{5}$O$_{12}$) film is confirmed with high-resolution X-ray diffraction of the GSGG (444) substrate and film peak, labeled respectively; see Extended Data Fig.~\ref{SI-Char}a.
The sample thickness was measured by X-ray reflectivity and the fitted thickness was $(21.7 \pm 0.5)$\,nm. An in-plane hysteresis loop was measured by vibrating sample magnetometry (VSM) yielding a saturation magnetization of $M_\mathrm{s}=(86\pm2)\,\mathrm{kA m}^{-1}$.

To estimate the anisotropy and the exchange constant, we measured micro-focused BLS on thermally excited spin waves in varying in-plane external magnetic fields. The obtained data were fitted using an analytical model under the assumption of unpinned spins at the surfaces. The first step is to calculate the static magnetization angle ($\theta$, measured from in-plane direction) by minimizing the total system energy, which is expressed as

\begin{multline}
\frac{E_{\text{tot}}}{V}
= \frac{E_{\text{ani}}}{V} + \frac{E_{\text{demag}}}{V} + \frac{E_{\text{Zeeman}}}{V}
= -K_{1} \sin^{2}(\theta) \\
- \frac{\mu_{0}}{2} M_{s}^{2} \cos^{2}(\theta)
- B_{\text{ext}} M_{s} \cos(\theta).
\end{multline}

The second step is to calculate the mode frequency

\begin{multline}
f^{2} = \left( \frac{\gamma}{2\pi} \right)^{2}
\Biggl(
    B_{\text{ext}} \cos\theta
    + \left( \frac{2K_{1}}{M_\mathrm{s}} - \mu_{0} M_\mathrm{s} \right) \\
    \cos^{2}\!\left(\tfrac{\pi}{2}-\theta\right)
    + \frac{2A_{\text{ex}}}{M_\mathrm{s}} \left(\frac{\pi n}{d}\right)^{2}
\Biggr) \\
\Biggl(
    B_{\text{ext}} \cos\theta
    + \left( \frac{2K_{1}}{M_\mathrm{s}} - \mu_{0} M_\mathrm{s} \right) \\
    \cos\!\bigl(2(\tfrac{\pi}{2}-\theta)\bigr)
    + \frac{2A_{\text{ex}}}{M_\mathrm{s}} \left(\frac{\pi n}{d}\right)^{2}
\Biggr).
\end{multline}

where $K_1$ is the total out-of-plane anisotropy constant, $n$ stands for the number of nodes in the out-of-plane direction (we fitted $n=0$ and $n=1$, see Extended Data Fig.~\ref{SI-Char}), $A_\mathrm{ex}$ is the exchange constant, $d$ is the film thickness, and $\gamma$ is the gyromagnetic ratio. The fit yields $K_1=(16780 \pm 60)\,\mathrm{J}\,\mathrm{m}^{-3}$ (due to magnetoelastic and magnetotaxial anisotropy \cite{kaczmarek2024atomic}) and $A_\mathrm{ex}=(2.847 \pm 0.003)\,\mathrm{pJ}\,\mathrm{m}^{-1}$.

The thickness of the Ga:YIG (approx. Ga$_{1.1}$Y$_{3}$Fe$_{3.9}$O$_{12}$) was measured using a prism coupler to be $(161 \pm 2)$\,nm. The saturation magnetization was estimated from the out-of-plane VSM loop as $M_\mathrm{s}=(15.5 \pm 0.1)\,\mathrm{kA\,m}^{-1}$, see Extended Data Fig.~\ref{SI-GaYIGChar}. The total anisotropy constant and exchange constant were estimated from STXM measurements to be $K_1=(1020 \pm 50)\,\mathrm{J\,m}^{-3}$ and $A_\mathrm{ex}= (1.90 \pm 0.05)\,\mathrm{pJ\,m}^{-1}$.

\subsection*{Micromagnetic simulations}
All micromagnetic simulations were performed in MuMax3 \cite{Vansteenkiste2014design}. The simulation world was set to $2048\times64\times4$ cells with sizes $16\times16\times5.425$\,nm$^3$, resulting in a size of approx. $33\times1\times0.02$\,µm$^3$ for Bi:YIG and $4096\times128\times16$ cells with sizes approx. $8\times8\times10$\,nm$^3$, resulting in a size of $32\times1\times0.161$\,µm$^3$ for Ga:YIG. The periodic boundary conditions were set in both in-plane directions with 4 and 128 repetitions for Bi:YIG and 2 and 64 repetitions for Ga:YIG to compensate for different sizes. The static external field was set in \textit{y}-direction. The excitation field for the dispersion relation was in the out-of-plane direction and was set as a 2D sinc pulse $B_\mathrm{ext} (x,t)=A \mathrm{sinc}(k_\mathrm{c} (x-x_0 )) \mathrm{sinc}(\omega_\mathrm{c} (t-t_0 ))$, where $k_\mathrm{c}=150$\,rad µm$^{-1}$ is the cut-off wavenumber, $\omega_\mathrm{c}=2\pi \times 60~\mathrm{GHz}$ is the cut-off frequency, $x_0,y_0$ are the centers of the simulation area, $t_0=100$\,ps is the time of maximum amplitude of the pulse, and $A=1$\,mT is the excitation amplitude. For simulations including the driving field from the antenna, we calculated this field with the analytic formula \cite{Devolder2023}. The out-of-plane magnetization component was saved every 8\,ps for the topmost layer. The simulation code and Python tools are available from \cite{zenodo}.

\subsection*{Time-resolved scanning transmission X-ray microscopy}
The time-resolved STXM measurements for imaging the dynamics of the out-of-plane component of magnetization were carried out at the MAXYMUS end station at the BESSY II electron storage ring operated by the Helmholtz-Zentrum Berlin für Materialien und Energie \cite{Weigand2022}. This method utilizes the specific time structure of the incident X-rays, which comprises pulses with a 2\,ns repetition rate and approx. 100\,ps effective pulse length. In STXM, a monochromatic X-ray beam is focused onto the sample by means of a diffractive zone plate. The locally transmitted X-ray intensity is measured by a single-pixel detector, so raster scanning the sample provides a 1D absorption image, where the dimension depicts the distance from the antenna, with a lateral step of approx. 200\,nm. Using circularly polarized X-rays allows X-ray magnetic circular dichroism to be exploited, thus obtaining a magnetic contrast \cite{Mayr2021, wintz2016magnetic}.

\subsection*{Time-resolved Brillouin light scattering microscopy}
We used a custom-developed optical setup to measure Brillouin light scattering spectra \cite{wojewoda2023observing, wojewoda2023phase, wojewoda2024modeling}. A single-mode laser (COBOLT Samba) with a wavelength of 532\,nm was used as a light source. The spectral purity of the laser light was improved by a Fabry-Perot filter (TCF-2, Table Stable). The incident power on the sample was 3\,mW, which did not introduce any visible nonlinear phenomena or heating of the sample. An optical microscope with active stabilization was used to compensate the mechanical drift of the sample (THATec Innovation). The light was focused and collected through the same objective (Zeiss LD EC Epiplan-Neofluar 100 × /0.75 BD). The inelastic frequency shift was measured with a tandem Fabry-Perot interferometer (TFP-2HC interferometer, Table Stable) and the time delay was measured by a Swabian Instruments Time Tagger 20. To generate the magnetic field, we used a water-cooled GMW 5403 electromagnet powered by two KEPCO BOP20-20DL power supplies and a predefined current-field calibration at the sample position.

\subsection*{Theoretical model and analysis}

For the theoretical analysis, we use the exchange model for spatio-temporal evolution of the dynamic magnetization variable $u(x,t)$:
\begin{equation}\label{e:th-init}
    u_t - iDu_{xx} + i(\omega_0 + T|u|^2)u + \Gamma u = f(x) e^{-i\omega_\mathrm{e} t}.
\end{equation}
Here $D = \omega_M \lambda^2$, $\lambda = \sqrt{2A_\mathrm{ex}/(\mu_0M_\mathrm{s}^2)}$ is an exchange length, $\omega_M = \gamma \mu_0 M_\mathrm{s}$, $\Gamma \approx \alpha_G \omega_0$ is the damping rate, $\omega_0$ is the FMR frequency. The nonlinear frequency shift is $T = \gamma^2(B_\mathrm{a} - \mu_0 M_\mathrm{s})|B_\mathrm{ext}|/(2\omega_0)$ with $B_\mathrm{a} = 2K_1/M_\mathrm{s}$ being the anisotropy field; under this convention the dynamic variable $u$ is related to the precession angle as $|u|^2 = 1-\overline{\cos\varphi}$ (overbar means time averaging) \cite{Krivosik_PRB2010}. For the Bi:YIG sample, we get $T/(2\pi) = 5.6\,$GHz. Finally, $f(x) = C\gamma b_\mathrm{RF}/2$ is the excitation force applied at the frequency $\omega_\mathrm{e}$ with the coefficient $C \approx 1$ determined by the precession ellipticity and polarization of the applied RF field. A stationary solution of Eq.~(\ref{e:th-init}) was used for the soliton profile plots in Extended Data Fig.~\ref{S-th}; somewhat higher values of the RF field were used to get a similar soliton size due to the neglected effect of dipolar interaction.

In the case $f(x) = 0$ (i.e. outside of the excitation region), it is possible to derive the following relation for the phase evolution of the stationary solution $u(x,t) = a(x)e^{i\psi(x)}e^{-i\omega_\mathrm{e} t}$:
\begin{equation}
    \frac{d\psi}{dx} = \frac{\Gamma}{D a^2(x)} \int_x^\infty a^2(\xi) d\xi,
\end{equation}
which means that it is impossible to realize a localized quasi-uniform solution ($d\psi/dx = 0$). Thus, a large-size soliton-like solution is stabilized by the spatially extended antenna near fields, as mentioned in the main text.

To understand the initial stage of magnetization precession onset, we neglect the nonlinear and dispersive terms in Eq.~(\ref{e:th-init}), arriving at a standard local damped oscillator model. The well-known solution with zero initial conditions at $t\to 0$ reads $u(t) \approx ft e^{-i\omega_\mathrm{e} t}$. For this reason, we use a linear fit when analyzing time-resolved data of soliton formation. The finite apparent delay of the magnetization onset in the experiment is a consequence of evolution from a random-phase thermal level.

\bibliography{sn-bibliography}

\clearpage

\renewcommand{\figurename}{Extended Data Fig.} 
\setcounter{figure}{0}                         
\renewcommand{\thefigure}{\arabic{figure}}     

\makeatletter
\renewcommand{\theHfigure}{ED.\arabic{figure}}
\makeatother

\providecommand{\figureautorefname}{Extended Data Fig.}
\renewcommand{\figureautorefname}{Extended Data Fig.}

\section*{Extended Data}\label{ExtData}
\renewcommand\figurename{Extended Data Fig.}%

\begin{figure*}
\centering
\includegraphics[width=1\textwidth]{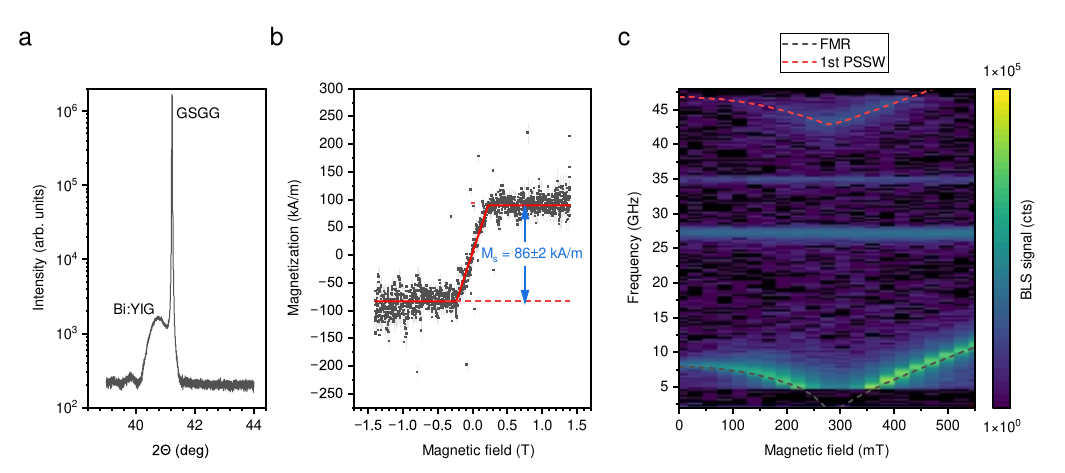}
\caption{\textbf{Characterization of the Bi:YIG sample}. \textbf{a}, X-ray diffraction data. \textbf{b}, Vibrating sample magnetometry hysteresis loop (squares) and three-line piecewise fit. The magnetic field was applied in-plane. \textbf{c}, BLS spectra on thermally excited spin waves in varying in-plane magnetic fields. The black ($n=0$) and red ($n=1$) dashed lines show fits to both modes; see Methods.}\label{SI-Char}
\end{figure*}

\begin{figure*}
\centering
\includegraphics{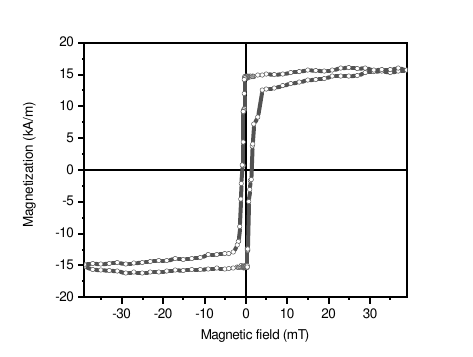}
\caption{\textbf{Characterization of the Ga:YIG sample}. Vibrating sample magnetometry hysteresis loop in out-of-plane magnetic field (circles). }\label{SI-GaYIGChar}
\end{figure*}

\begin{figure*}
\centering
\includegraphics[width=1\textwidth]{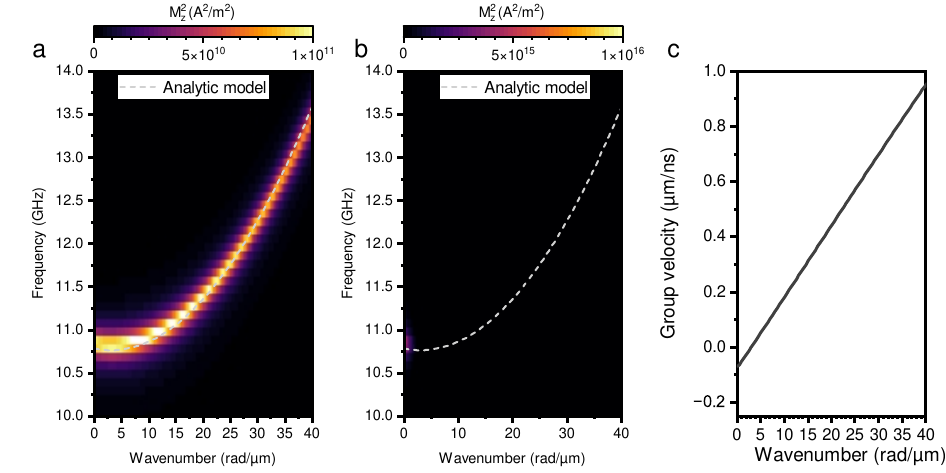}
\caption{\textbf{Linear dispersion relation and group velocity obtained from micromagnetic simulations and analytical modeling}. \textbf{a,b,} The 2D color code shows the distribution of the amplitude of the out-of-plane magnetization component. This quantity represents a Bloch function (density of the spin-wave states with respect to wavenumber and frequency). The magnetization dynamics was excited with a 2D sinc pulse \textbf{(a)} and with a magnetic field produced by the antenna \textbf{(b)}. The gray dashed line shows the results of an analytical calculation based on the linearized Landau–Lifshitz–Gilbert equation. \textbf{c}, Group velocity calculated from the analytical model.}\label{S-Dispersion}
\end{figure*}

\begin{figure*}
\centering
\includegraphics[width=16cm]{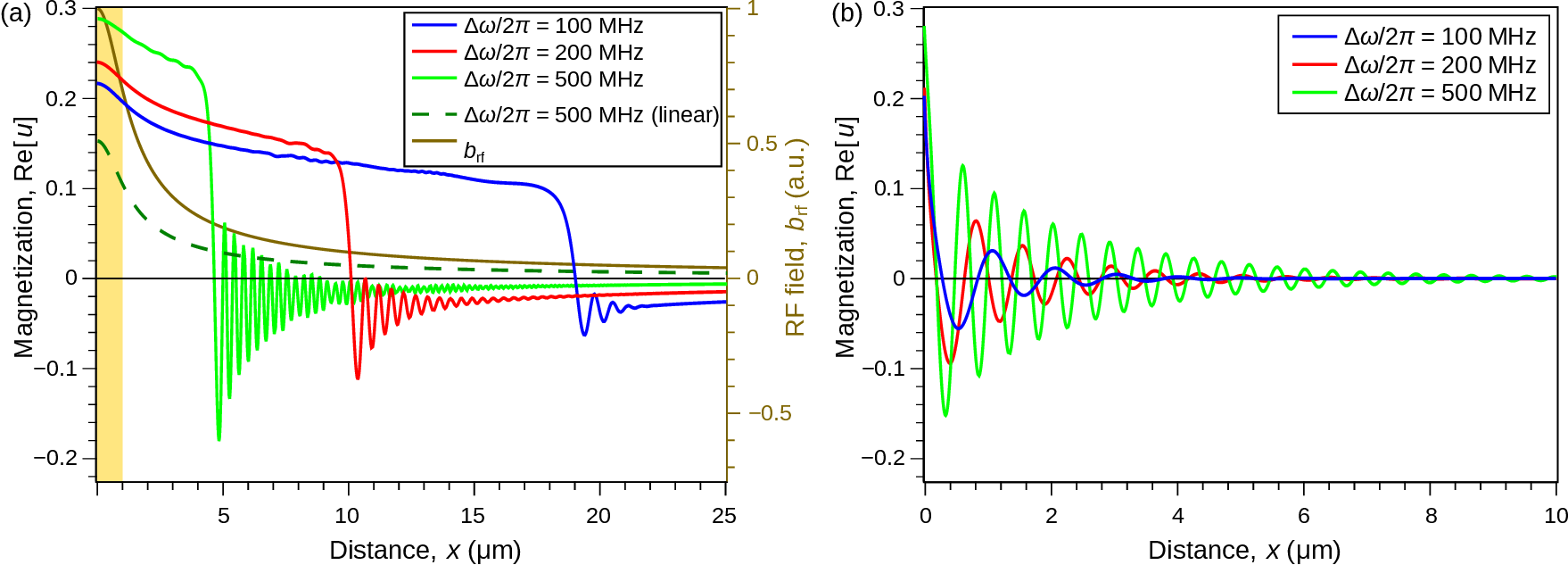}
\caption{\textbf{Analytical model.} \textbf{a,} Stationary soliton profiles (dimensionless dynamic magnetization $\mathrm{Re}[u]$) for different offsets $\Delta\omega$ of the driving field frequency from the FMR, accounting for the microstrip near fields; the excitation field profile is shown by the dark yellow curve (right axis), with the microstrip position shown by the shaded area; a half-space is shown, i.e. $x = 0$ corresponds to the microstrip center; maximal field amplitude $b_\mathrm{RF}(x=0) = 40\,$mT. The dashed line shows the magnetization profile in the linear case ($T = 0$), which is much more localized than the soliton; variation of $\Delta\omega$ changes only the amplitude, but not the profile. \textbf{b,} Profile of nonlinear spin-wave excitations when the antenna near field is neglected outside the antenna, showing no soliton formation; initial amplitudes at $x = 0$ are taken comparable to those in panel (a).}\label{S-th}
\end{figure*}

\begin{figure*}
\centering
\includegraphics{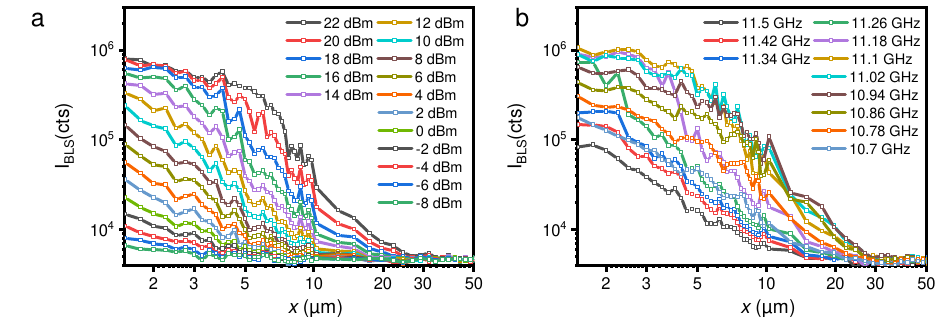}
\caption{\textbf{Power- and frequency-dependent BLS intensity measured on Bi:YIG.} \textbf{a} BLS signal dependence on the distance \textit{x} from the antenna at various driving power levels. With lower driving power, the spatial extent of the soliton decreases, while for powers below approx. 6\,dBm we cannot observe any soliton at all. \textbf{b,} BLS signal dependence on the distance \textit{x} from the antenna at various driving frequencies. For frequencies around the resonance (10.78 -- 11.34\,GHz), we can observe the formation of the soliton with a maximum extent around 11.02\,GHz. When out of resonance ($\leq10.7\,$GHz or $\geq11.42\,$GHz) we cannot observe the formation of any soliton.}\label{S-PowerFreq}
\end{figure*}

\begin{figure*}
\centering
\includegraphics{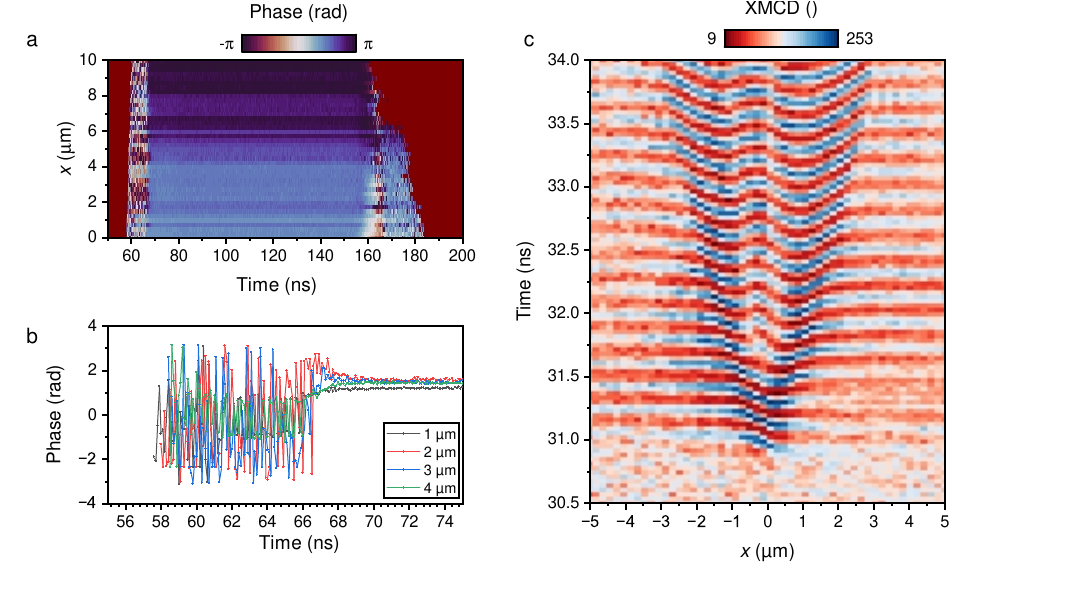}
\caption{\textbf{Phase evolution during soliton formation.} \textbf{a,} Phase- and time-resolved BLS measurements during the formation of the soliton, excited at 11.02\,GHz and 550\,mT for Bi:YIG. \textbf{b,} Line cuts extracted from panel \textbf{a}. During the amplitude build-up, the phase exhibits a nontrivial evolution, with the spatial slope even changing sign depending on the distance from the antenna. \textbf{c,} STXM measurement at a static field of 190\,mT for Ga:YIG. Before the soliton stabilizes around 0\,µm, the magnetization evolution exhibits a finite spatial slope.}\label{S-PR-TR}
\end{figure*}

\begin{figure*}
\centering
\includegraphics{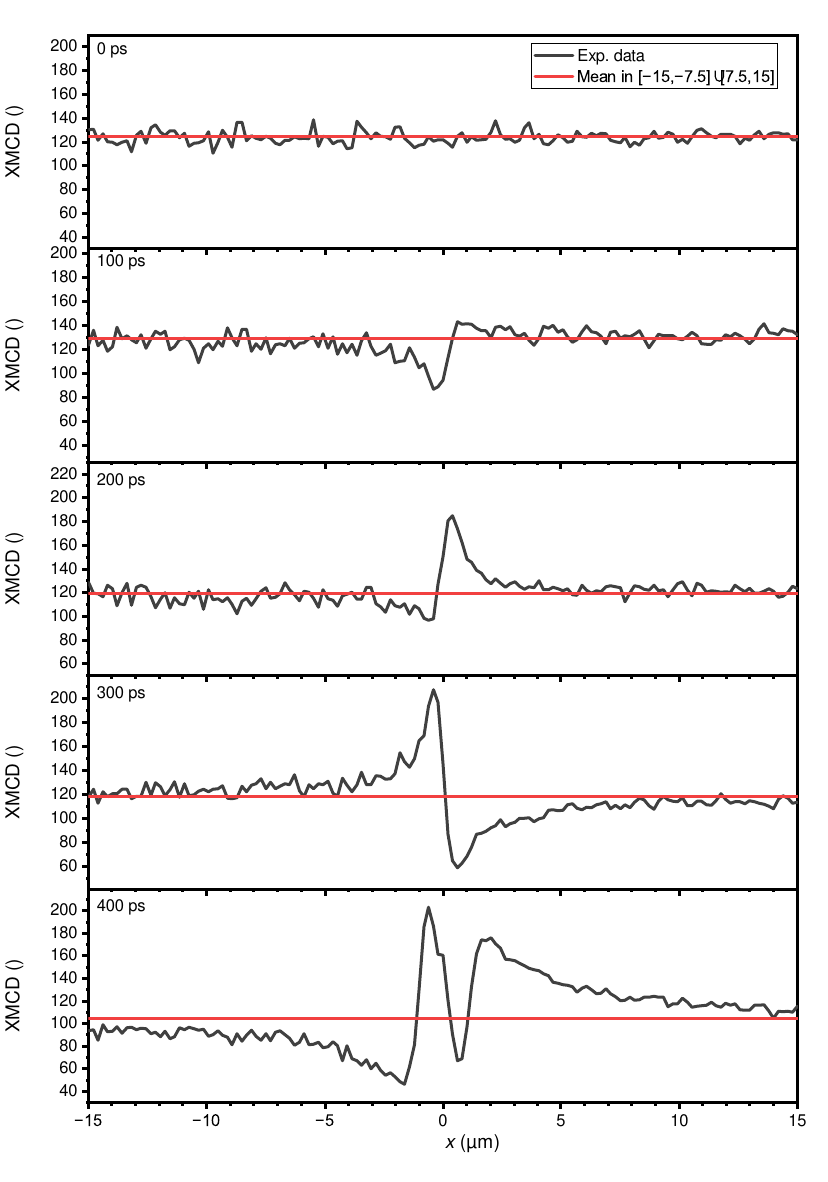}
\caption{\textbf{Time-resolved STXM data on Ga:YIG.} The black solid lines represent the experimental data measured by STXM with an excitation frequency of 3.75\,GHz. The red lines show their average over the distance interval from $-15\,$\textmu m to $-7.5\,$\textmu m and from $7.5\,$\textmu m to $15\,$\textmu m. The offset in intensity at negative and positive distances becomes clear after 200\,ps throughout the whole measured region. From this, one can estimate that the group velocity exceeds 75\,\textmu$\mathrm{m\,} \mathrm{ns}^{-1}$. At 300\,ps, this offset changes sign, which indicates temporal oscillation. After 400\,ps, the onset of the soliton formation is visible.}\label{S-GaYIG-XMCD}
\end{figure*}

\end{document}